\documentclass[11pt]{article}
\usepackage{moriond,epsfig}

\begin{document}
\vspace*{4cm}
\title{FERMI-DIRAC CORRELATIONS IN $Z^{0} \rightarrow ppX$ AT LEP}

\author{ M. KUCHARCZYK }

\address{Henryk Niewodniczanski Institute of Nuclear Physics, Polish Academy of Sciences,\\
ul. Radzikowskiego 152, 31-342 Cracow, Poland}

\maketitle\abstracts{
The DELPHI and ALEPH experiments have released new results of the Fermi-Dirac correlations for $p p$ ($\bar{p} \bar{p}$) and Bose-Einstein correlations for $K^{0}_{s} K^{0}_{s}$. Both experiments measure very small source radius for protons $R \sim 0.1$ $fm$. The source dimension for $K^{0}_{s} K^{0}_{s}$ is in agreement with the previous measurements for $K^{+} K^{-}$, $K^{0}_{s} K^{0}_{s}$. These results, together with earlier LEP measurements, establish the dependence of the correlation radius on the hadron mass. This paper discusses some of the attempts to describe this phenomenon.}

\section{Introduction}

The Fermi-Dirac (FDC) or Bose-Einstein correlations (BEC) are examined by measuring a two-particle correlation function defined as a ratio of the inclusive density distribution for two particles and the so-called reference density. The reference density is a two-particle density distribution which approximates the distribution without the FDC or BEC effect. The central problem in the FDC or BEC analyses is the definition of the reference sample. There are three different ways of defining the reference sample: (a) the Monte Carlo sample without the FDC or BEC, (b) an event-mixed sample based on the choice of two identical hadrons, each originating from a different data event, (c) the distribution for opposite charged particle pairs (the so-called "unlike" distribution). In the reported analyses the correlations are studied in the four-momentum difference $Q$, where the correlation function is fitted using a Goldhaber-like parametrisation \cite{gold}:
\begin{equation}
C(Q) = N(1 \pm \lambda e^{-Q^{2}R^{2}}),
\label{eq:gold}
\end{equation}
\noindent where $R$ is the source radius, $\lambda$ is the so-called chaocity parameter which characterises the strength of the correlations, $N$ denotes the normalisation factor. In the above equation the sign $"+"$ is assigned to the BEC effect, $"-"$ to the FDC effect.

\section{Fermi-Dirac correlations for protons (antiprotons)}

\subsection{DELPHI analysis for antiprotons$^{2}$} \label{subsec:delphi_pp}

The data sample of $\sim 2$ M of $Z^{0} \rightarrow hadrons$ events, recorded in 1994 and 1995 in the DELPHI detector at LEP was used. The Monte Carlo simulation sample was obtained using the JETSET 7.4 generator (without the FDC effect) and contained $\sim 6$ M hadronic events. Since it was crucial to identify the particles the requirement for the minimal particle momentum of $0.7$ $GeV$ was used. In this momentum region the MC agrees very well with the data and the identification information is available from both the RICH detector and the $dE/dx$ measurement. After applying the  selection cuts there remained about $180$ K events with at least two proton candidates for the analysis in data and about $560$ K in the MC sample. The single proton purity was $85\%$ at the efficiency of $70\%$, as determined from the MC simulation. This resulted in $70\%$ purity for the $\bar{p} \bar{p}$ sample and $50\%$ purity for the $p p$ sample. The cleaner $\bar{p} \bar{p}$ sample is used for the source radius measurement. The analysis of the $p p$ sample, after background subtraction, yielded consistent results. The correlation function was determined using three different reference samples: the Monte Carlo, event-mixed and unlike sign pairs. The correlation function, together with fit results based on parametrisation (\ref{eq:gold}) for $\bar{p} \bar{p}$, constructed with respect to three different reference samples, are shown in Fig.\ref{delphi_pp}. Since the analysis relying on the event-mixing reference sample was more precise than the MC or unlike analyses, the final result was taken from the event-mixing analysis: $R = 0.16 \pm 0.04(stat) \pm 0.03(syst)$ $fm$, $\lambda = 0.67^{+0.19}_{-0.17}(stat) \pm 0.18(syst)$. The other two methods were used to estimate the systematic error. The main contributions to the systematic error come from the proton identification, variations in different reference samples, parametrisation of the correlation function and binning width in $Q$.
{\bf
\begin{figure}[h]\centering
\mbox{\epsfxsize 12.0cm\epsfysize 9.5cm\epsfbox{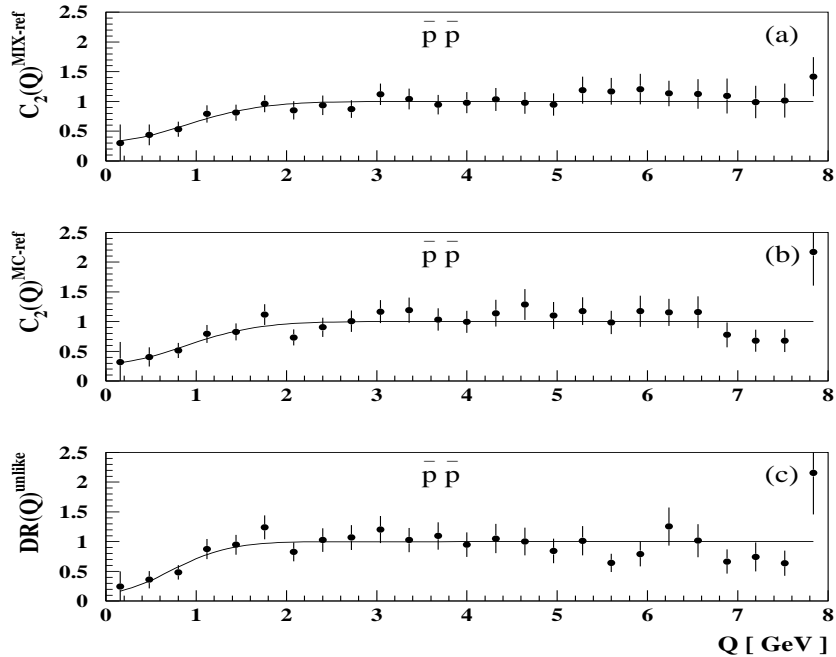}}\\
\caption{Correlation function for $\bar{p} \bar{p}$ (DELPHI preliminary results): (a) event-mixed reference sample, (b) MC reference sample, (c) 'unlike' sample. Solid lines denote fit results using parametrisation (\ref{eq:gold}).}
\label{delphi_pp}
\end{figure}
}

\subsection{ALEPH analysis for protons (antiprotons)$^{3}$}\label{subsec:aleph_pp}

\begin{figure}[h]
\hbox{
\vspace{1.0cm}
\epsfxsize=80mm
\epsfysize=50mm
\epsfbox[43 150 536 528]{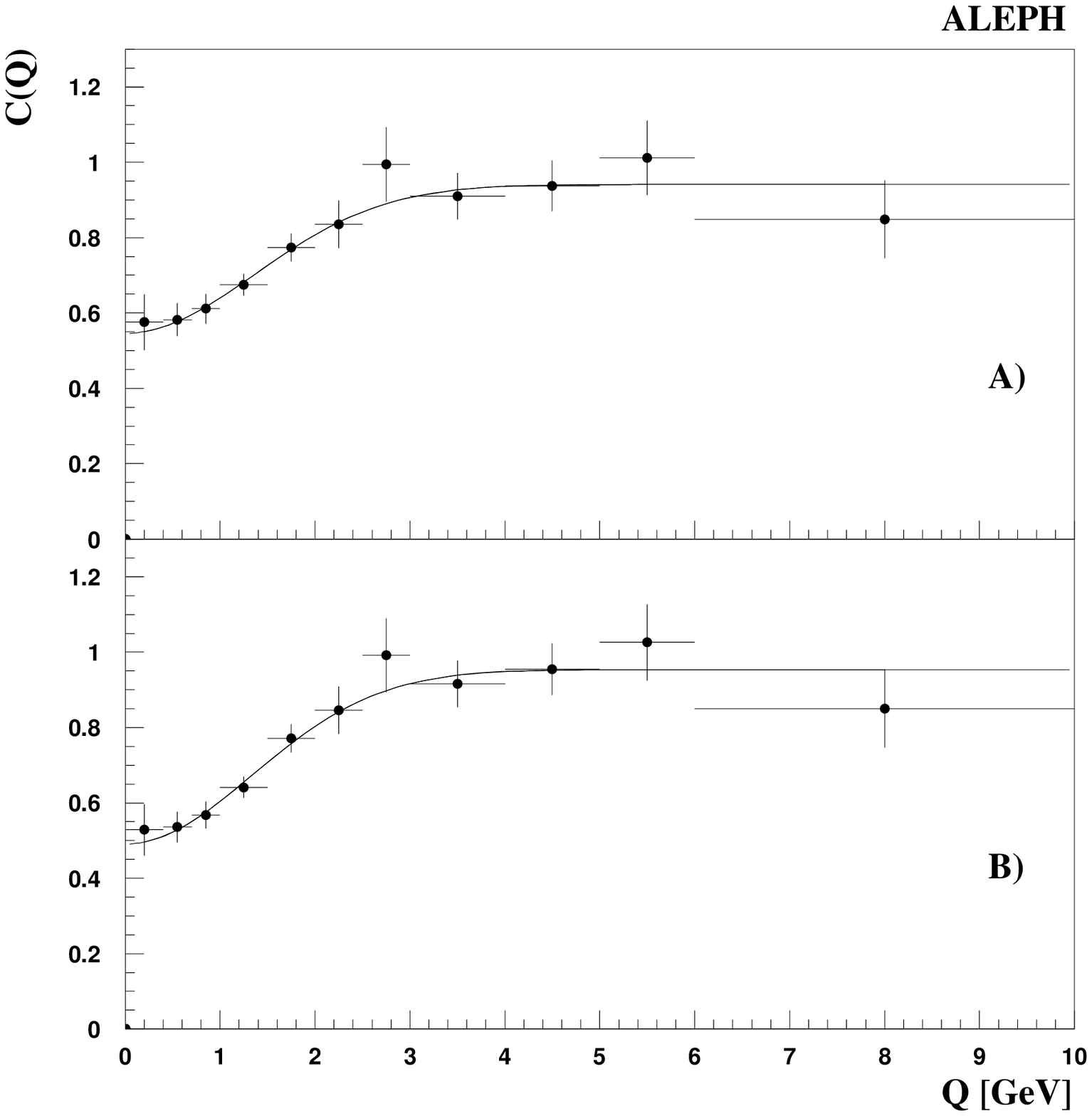}
\epsfxsize=80mm
\epsfysize=50mm
\epsfbox[43 150 536 528]{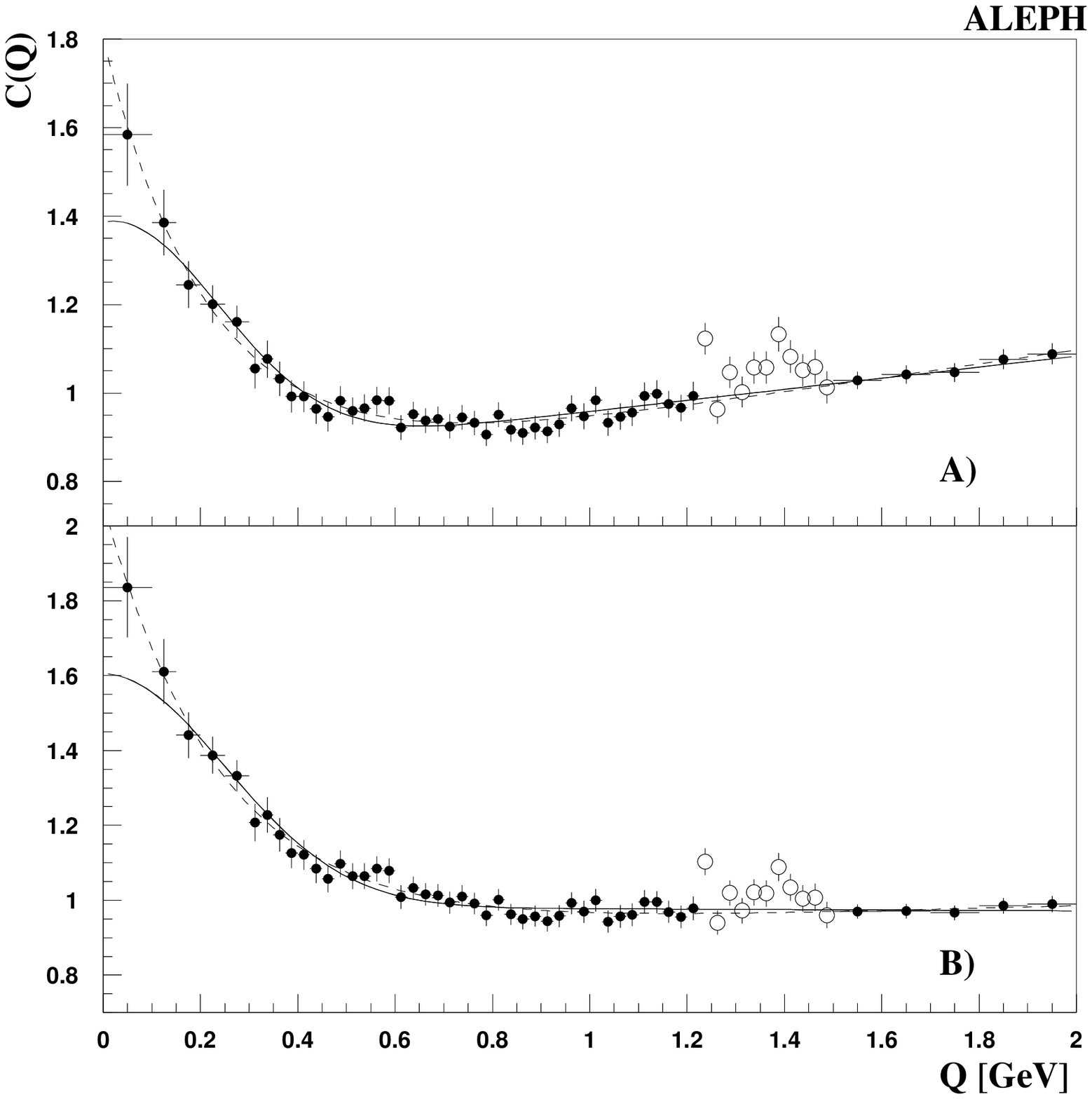}}
\vspace{0.5cm}
\caption{ ALEPH results (preliminary): {\bf (Left)} $C_{2}(Q)$ for $p p$ and $\bar{p} \bar{p}$ pairs using different reference samples: (A) Monte Carlo, (B) mixed event double ratio. The solid lines indicate the results of the fits using the parametrisation (\ref{eq:gold}). {\bf (Right)} $C_{2}(Q)$ for $K^{0}_{s} K^{0}_{s}$ pairs using different reference samples: (A) Monte Carlo, (B) mixed event double ratio. The curves represent the results of the fits using parametrisation (\ref{eq:gold}) (solid lines) and the exponential parametrisation (dashed lines). The region of $f_{0}(1710)$ (open circles) has been excluded from the fits.}  
\label{aleph}
\end{figure}

The analysed data sample consisted of a total of $3.9$ M $Z^{0} \rightarrow hadrons$ events and $6.5$ M Monte Carlo events obtained employing the JETSET 7.4 generator (without the Fermi-Dirac correlation effect). After applying the selection cuts the purity for a single proton was $70\%$, which gave a $74\%$ purity for (anti)proton-(anti)proton pairs for $Q < 5.0$ $GeV$. The fits to the correlation function are shown in Fig.\ref{aleph}(left). The discrepancy between the proton momentum distributions in the data and the MC simulation in the region $0.5 < p < 1.3$ $GeV$ was corrected by using momentum dependent weights. The region $1.3 < p < 3.0$ $GeV$ was excluded from the analysis since the identification was based only on $dE/dx$. The analysis was conducted for the event-mixed and the Monte Carlo reference samples for both $p p$ and $\bar{p} \bar{p}$ pairs. The final result quoted for the event-mixing analysis is: $R = 0.10 \pm 0.02(stat) \pm 0.02(syst)$ $fm$, $\lambda = 0.46 \pm 0.04(stat) \pm 0.11(syst)$. The main contributions to the systematic error are due to: different reference samples, the Monte Carlo reweighing, the Coulomb corrections.

\section{ALEPH analysis of Bose-Einstein correlations for $K^{0}_{s}K^{0}_{s}$$^{3}$}

The analysis was based on the sample of $\sim 200$ K events with two $K^{0}_{s}$. The purity of the selected $K^{0}_{s}K^{0}_{s}$ sample was $96\%$, the efficiency $27\%$. The correlation function was obtained using two reference samples: the Monte Carlo and event-mixed. The fits to the correlation function were performed with two different parametrisations: the Goldhaber-like (\ref{eq:gold}) and the exponential (Fig.\ref{aleph}(right)). The long range correlations had to be taken into account for the Monte Carlo reference sample in the region of $Q>0.8$ $GeV$. The presence of the $f_{0}(980)$ resonance, simulated in the MC with the width of $0.1$ $GeV$, affects the first bin. The $Q$ region corresponding to the $f_{0}(1710)$ resonance was excluded from the fit. The above effects account for the systematic error. The final result comes from the event-mixing analysis: $R = 0.57 \pm 0.04(stat) \pm 0.06(syst)$ $fm$, $\lambda = 0.63 \pm 0.06(stat) \pm 0.06(syst)$.

\section{Correlation radius dependence on hadron mass}

The above results, together with the earlier LEP results \cite{lep}, establish a strong decrease of $R$ with the increase in hadron mass. There are three approaches to understand this effect: semi-classical derivations based on the Heisenberg relations or the virial theorem \cite{alex} and a quantum-mechanical model based on the Bj\"orken-Gottfried condition \cite{bjorken}.

The first attempt \cite{alex} links the $R(m)$ behaviour to the Heisenberg uncertainty principle. Using the Heisenberg relations in momentum-space and energy-time, and identifying $\Delta t$ with the time scale of strong interactions ($\Delta t = 10^{-24}$ $s$, independent of the hadron mass), one obtains $R(m) = \frac{0.243}{\sqrt{m}}$, which correctly reproduces the data.

In the second attempt \cite{alex}, the virial theorem is used to estimate the relative momentum of two particles bound by the potential $V(R)$. For the general QCD potential $V(R) = \kappa R - \frac{4}{3} \frac{\alpha_{s} \hbar c}{R}$ with the parameters set to the values obtained from the hadron wave functions and decay constants, the radius of the particle emission region changes according to the formula: $R \sim 1 / \sqrt{m}$.

Both explanations suggest that hadrons are emitted not from a unique source, but from the sources with radii strongly dependent on the mass. This makes the energy density of the source emitting protons or $\Lambda^{0}$'s unacceptable \cite{alex}.

In the third attempt to understand $R(m)$ \cite{bz}, the radius of the particle source is universal. The observed $R(m)$ dependence is solely the consequence of the Bj\"orken-Gottfried relation, which connects the space-time position $x_{\mu}$ of an emitted hadron to its four-momentum $q_{\mu}$: $q_{\mu} = \lambda x_{\mu}$. This relation is used in a quantum-mechanical formulation of the model by postulating that the source function is factorised: $S(P,X) = F(\tau) S_{\|} S_{\bot}$, with the $S_{||}$ and $S_{\bot}$ having Gaussian forms representing space-momentum correlations. The source function is related to the density matrix in momentum space by the Fourier transform $\rho(q,q') = \int{d^{4} X e^{iQX} S(P,X)}$, where $P = (q + q') / 2$, $Q = q - q'$ and $q$, $q'$ are four-momenta of a particle. Therefore, if the source function is known, the one- and two-particle distributions are determined completely. The parameters of the model: the ranges of position-momentum correlations, were constrained by the fit of the predicted $p_{T}$ distribution to the data \cite{bkpz}. The model reproduces correctly the observed $R(m)$ dependence and the longitudinal source elongation. An important consequence of the model is the fact that the observed source radius is the apparent radius, which is smaller than the real particle emission radius:
\begin{equation}
\frac{1}{R^{2}_{obs}} \approx \frac{1}{R^{2}} + \frac{M^{2}_{\bot}}{\tau^{2}_{0}\delta^{2}} \qquad \Rightarrow R_{obs} < R,
\label{radius}
\end{equation}
\noindent where $\tau_{0}$ is the longitudinal proper time, $\delta$ parametrises the correlation range.


\section{References}
\begin{thebibliography}{99}

\bibitem{gold}
G. Goldhaber et al., Phys. Rev. {\bf 120} (1960) 300.\\
G. Goldhaber et al., Phys. Rev. Lett. {\bf 3} (1959) 181.

\bibitem{delphi_ap}
M. Kucharczyk, H. Palka, CERN-EP/2004-003, DELPHI-CONF-684 (2004).

\bibitem{aleph}
ALEPH Conference Note.

\bibitem{lep}
G. Alexander, Rept. Prog. Phys. 66 (2003) 481-522.

\bibitem{alex}
G. Alexander, I. Cohen, E. Levin, Phys. Lett. {\bf B452} (1999) 159.

\bibitem{bjorken}
K. Gottfried, Phys. Rev. Lett. {\bf 32} (1974) 957.

\bibitem{bz}
A. Bialas, K. Zalewski, Acta. Phys. Polon. {\bf B30} (1999) 359.

\bibitem{bkpz}
A. Bialas, M. Kucharczyk, H. Palka, K. Zalewski, Phys. Rev. {\bf D62} (2000) 114007.\\
A. Bialas, M. Kucharczyk, H. Palka, K. Zalewski, Acta Phys. Polon., {\bf B32} (2001) 2901.



\end{thebibliography}
\end{document}